\documentstyle[12pt]{article}
\begin {document}
\newcommand {\beqa}{\begin{eqnarray}}
\newcommand {\eeqa}{\end{eqnarray}}
\newcommand {\n}{\nonumber \\}
\newcommand {\beq}{\begin{equation}}
\newcommand {\eeq}{\end{equation}}
\newcommand {\om}{\omega}
\newcommand {\s}{\sigma}

\newcommand {\pa}{\partial}
\newcommand {\e}{\epsilon}
\newcommand {\th}{i\theta}
\newcommand {\ga}{\gamma}
\begin{flushright}
RI-147
\\
July  1992
\end{flushright}
\begin{center}
{\large\bf Deviations from Scale Invariance  \\
      near \\
      a General Conformal Background} \vspace{.7 in}\\
{\bf A. Babichenko and
        S. Elitzur\footnote{Supported by the Israeli Academy for Sciences}
        \vspace{.5 in}\\
        {\it Racah institute of Physics, Hebrew University }\\
        {\it Jerusalem, 91904, Israel \vspace{.8 in}}}
\end{center}
\begin{abstract}
  Deviations from scale invariance resulting from small perturbations
of a general two dimensional conformal field theory are studied.
They are expressed in terms of beta functions for renormalization of
general couplings under local change of scale. The beta functions for
homogeneous background are given perturbatively in terms of the data of the
original conformal theory without any specific assumptions on its nature.
The renormalization of couplings to primary operators and to first descendents
 is considered as well as that of couplings of a dilatonic type
which involve explicit dependence on world sheet curvature.
\end{abstract}
\newpage

{}~
\section{Introduction}

   The condition for a consistent background for motion of a critical string
is that the corresponding world sheet model should be a conformal field theory
with a specified total anomaly. This means that the theory is a fixed point of
the renormalization group with vanishing $\beta$-functions. For a string
moving in $d$-dimensional target space this requirement was shown long ago to
imply Einstein equation on the geometry of target space to leading order
in the deviation from a flat space \cite{1}.
It was also found \cite{2}, that the target space field corresponding to the
dilaton particle is expressed in this context as an additional coupling in
the world sheet theory which involves explicitly the world sheet curvature.
This type of coupling is induced by renormalization and through the requirement
 of vanishing of $\beta$-function also the dilaton field acquires its
appropriate
 target space dynamics.

   A flat $d$-dimensional target space is just a very special case of a
conformal field theory. Although there is no general classification of all
these theories,there is a general framework which assigns to each conformal
theory some characterizing data like its central charge,chiral algebra,
spectrum of primary fields and operator product coefficients \cite{3},
\cite{4}. Each such a theory can consist a part of a consistent background
for motion of a string. It is natural then to generalize the $\sigma$-model
approach described above and ask for the deviation from scale invariance
resulting from small general perturbation around any conformal field theory.
This deviation should involve $\beta$-functions expressed universally in
terms of the conformal data of the perturbed theory without reference to
its specific nature.

   Concentrating on the flow of the coupling to the Virasoro primary fields,
which are most relevant, these $\beta$-functions were written to second order
in the perturbations by Zamolodchikov \cite{5} and Cardy and Ludwig
\cite{6}. They found the simple expression

\beq
\beta^i~=~\e_i g^i+\sum_{j,k} C^{i}_{jk} g^j g^k \label{1}
\eeq
where $g^i$ is the coupling corresponding to a primary operator of dimension
$2-\e_i$. For almost marginal perturbation eq. (\ref{1}) can be used to find
approximate new fixed points near the original one \cite{5},\cite{6}.

   The $\s$-model renormalization equations of ref \cite{1} should actually
be a special application of eq. (\ref{1}) which deals with a general
conformal field theory.
 However, since eq. (\ref{1}) stops at the level of primary fields the full
structure of the sigma model is not revealed, in particular dilaton couplings
are not included. It is the purpose of this work to extend eq. (\ref{1}) to
the first descendent level and to take into account operators which are
explicitly sensitive to the world sheet metric being proportional to its
curvature. We will still ignore higher level descendents and higher powers
 and derivatives of the world sheet curvature, which are also induced by the
renormalization flow in a form dictated by the particular regularization
scheme being used, but are less relevant. Since we do not commit ourselves
to any special property of the conformal theory perturbed around, we cannot
use regularizations which depend on such properties, e.g. schemes that rely
on Feynmann integrals. Rather, following \cite{5} and \cite{6}, we will use
the straightforward and intuitive cut off scheme which does not allow
interaction vertices to get too close to each other on the world sheet.

   Unlike \cite{5} and \cite{6} we need a renormalization scheme which is
sensitive to the world sheet geometry. We will, therefore, adopt the point of
view that the renormalization flow is induced by a local conformal change
of the metric rather than by a global change in cut off scale \cite{7,8}.
In other words we are actually calculating the trace of the energy momentum
tensor of the perturbed model, including quantum corrections. The dependence
of the action on the background metric is dictated by its required general
covariance. Of course, once the change of scale is made local, the responding
renormalization of coupling constant is local as well resulting in a theory
with couplings which vary from point to point on the world sheet. Also
operators of nonzero spin are induced which couple to derivatives of the
curvature. We will choose our perturbed background to include only constant
coupling on a flat world sheet geometry, and study the infinitesimal
renormalization response of this background to local change of scale. For the
purpose of identifying new approximate conformal points, for which this
response
 vanishes, that should be enough, since at a conformal point there is no
dependence on the background world sheet metric and it can always be brought
into a flat form. We will follow the renormalization of primary fields and
of first level descendents including dilatonic interactions proportional to
the scalar world sheet curvature. Higher powers and derivatives of the
curvature will not be followed being less relevant. Since no derivatives of
the curvature are involved, only scalar operators, of total spin 0, are
induced.

   The kind of analysis done here has also bearing on a more ambitious
problem. Suppose we try to couple the non conformal perturbed model to quantum
two dimensional gravity. Then one should integrate over all local scales
where for each metric the matter action is the running one, the solution of
the local renormalization group. So just to formulate the action of the matter
plus gravity, the local renormalization group including all the metric
dependent
 terms should be solved. Of course our crude approximation including just the
most relevant terms near flat geometry is still very far from that goal.

   In the next section a general conformal model perturbed by primary fields
will be studied. We will show that renormalization induces changes in the
coupling constants as well as new perturbing non primary operators and
dilatonic
 coupling. Then in section 3 we will include such perturbations in the original
 action and calculate the full renormalization flow to that level.

{}~

\section{Primary Perturbations}

  Let $S_0$ be the action of some conformal field theory. Let
$O_{i}(z,\bar{z})$
 be primary fields within the theory with a respect to both left and right hand
 side Virasoro algebra of dimensions $h_i$ and $\bar{h}_{i}$ respectively.
Besides the spectrum of dimensions the theory is characterized by operator
product coefficients $C^{k}_{ij}$, namely for any quantity $A$  and operators
$O_i$ and $O_j$, the expectation value in the $S_0$ theory of
$<O_{i}(z,\bar{z})O_{j}(\om,\bar{\om})A>_0$ satisfies the expansion

\beqa
&&\left. <O_{i}(z,\bar{z})O_{j}(\om,\bar{\om})A>_0 =\sum_{k} C^{k}_{ij}
(z-\om)^
{h_k-h_i-h_j}(\bar{z}-\bar{\om})^{\bar{h}_k-\bar{h}_i-\bar{h}_j}
\right[ <O_{k}(\om,\bar{\om})A>_0
 \label{2} \\
&&\left. +\sum_{\{m\},\{\bar{m}\}} \ga^{(ijk)}_
{\{m\}} \ga^{(ijk)}_{\{\bar{m}\}} (z-\om)^{|m|} (\bar{z}-\bar{\om})^{|\bar{m}|}
<L_{-\{m\}}\bar{L}_{-\{\bar{m}\}} O_{k}(\om,\bar{\om})A>_{0}\right] \nonumber
\eeqa
\newline
Here $\{m\}$ stands for a series of positive integers $m_1,m_2,\ldots,m_n$,
$|m|=\sum_{i} m_i$ and $L_{-\{m\}} =L_{-m_{1}}L_{-m_{2}}... L_{-m_{n}}$.
The coefficients $\ga^{(ijk)}_{m}$ are determined algebraicly in terms of
$h_i$,$h_j$ and $h_k$ by the Virasoro algebra \cite{3}. In particular
$\ga^{(ijk)}_{1}$ multiplying the operator $L_{-1}O_{k}=\pa O_k$ turns out
to be

\beq
\ga^{(ijk)}_{1}=\frac{h_{i}-h_{j}+h_{k}}{2h_{k}} \label{3}
\eeq
\newline
   Let us perturb the conformal theory by adding to its action some small
combination of its primary operators

\beq
S=S_0 + \int d^{2}z \sum_{i} a^{-\e_{i}} g^{i}O_{i}(z,\bar{z}) \label{4}
\eeq
\newline
Here $\e_{i}=2-h_{i}-\bar{h}_{i}$, $a$ is some short distance cut off scale
which is needed to define the theory since after the perturbation it is
not conformal any more, and $g^{i}$ is small dimensionless coupling constants.
Eq. (\ref{4}) defines the theory on the flat background metric. Since after
perturbation the theory is not conformal its formulation depends also on the
world sheet metric. Choose the conformal gauge

\beq
ds^{2} = e^{2\s (z,\bar{z})} dzd\bar{z} \label{5}
\eeq
Note that in this gauge the only nonvanishing components of the connection are

\beq
\Gamma^{z}_{zz} = 2\pa\s ~~~~~~ \Gamma^{\bar{z}}_{\bar{z}\bar{z}} =
2\bar{\pa}\s \label{6}
\eeq
and the scalar curvature is

\beq
R = -e^{-2\s }\pa\bar{\pa}\s \label{7}
\eeq
where we put $\pa =\frac{\pa }{\pa z},~\bar{\pa} =\frac{\pa }{\pa \bar{z}}$.

  For general covariance the perturbing Lagrangian density should transform as
a scalar density. The operator $O_{i}$ of dimension $(h_{i},\bar{h}_{i})$ is
actually a tensor with $h_{i}$ indices of type $z$ and $\bar{h}_{i}$ of type
$\bar{z}$. To construct a scalar we must have $h_{i}=\bar{h}_{i}$ i.e. $O_{i}$
should have spin $S_{i}=h_{i}-\bar{h}_{i}=0$, and then contract the $h_{i}$
indices by $(g^{z\bar{z}})^{h_{i}}= e ^{-2h_{i}\s }$. Further multiplying by
$\sqrt{g} = e^{2\s }$ to give a density, the perturbation on a general
background metric becomes

\beq
S=S_0 + \int d^{2}z \sum_{i} a^{-\e_{i}} e^{\e_{i}\s (z,\bar{z})}
g^{i}O_{i}(z,\bar{z}) \label{8}
\eeq
\newline
Let $A$ be an arbitrary operator and$<A>$ its expectation value in the
perturbed theory. Up to second order in the couplings we have

\beqa
&& <A>=<A>_{0} + \sum_{i}~a^{-\e_{i}} g_{i} \int d^{2}z~
e^{\textstyle\e_{i}\s (z,\bar{z})} <O_{i}(z,\bar{z})A>_{0} + \label{9} \\
&& +\frac{1}{2}\sum_{ij}~a^{\textstyle -(\e_{i}+\e_{j})}g^{i}g^{j}
\int d^{2}zd^{2}\om ~
e^{\textstyle \e_{i}\s (z,\bar{z})+\e_{j}\s (\om,\bar{\om})}\times \n
&&\times <O_{i}(z,\bar{z})O_{j}(\om,\bar{\om})A>_{0}\theta(d(z,\om )-a)
\nonumber
\eeqa
\newline
The $\theta$-function in (\ref{9}) contains our coordinate independent
regularization prescription which avoids the singularity in $<O_{i}(z)
O_{j}(\om)A>$ at the coincidence of $z$ and $\om$ by demanding that
$d(z,\om)$, the geodesic distance between the points $z$ and $\om$ in the
metric (\ref{5}) is larger than the cut off $a$. Under an arbitrary small
variation of the conformal factor of the metric(\ref{5}), $\s \rightarrow\s +
\delta\s $, the variation of $<A>$ is

\beqa
&&\delta<A>=\frac{c}{6\pi}\int d^{2}z~e^{2\s} R(z,\bar{z})<A>_{0}
\delta\s (z,\bar{z})+ \label{10} \\
&&+\sum_{i}~\int d^{2}z~
G(z,\e_{i})\e_{i}g^{i}<O_{i}(z,\bar{z})A>_{0}
\delta\s (z,\bar{z})+ \n
&&+\frac{1}{2}\sum_{ij}g^{i}g^{j}\int d^{2}zd^{2}\om ~
G(z,\e_{i})G(\om,\e_{j})
(\e_{i}\delta\s (z,\bar{z})+\e_{j}\delta\s (\om,\bar{\om}))\times \n
&&\times<O_{i}(z,\bar{z})O_{j}(\om,\bar{\om})A>_{0}\theta(d(z,\om )-a)+ \n
&&+\frac{1}{2}\sum_{ij}g^{i}g^{j}\int d^{2}zd^{2}\om ~
G(z,\e_{i})G(\om,\e_{j})
<O_{i}(z,\bar{z})O_{j}(\om,\bar{\om})A>_{0}\times \n
&&\times \delta(d(z,\om)-a)\delta d(z,\om) \nonumber
\eeqa
\newline
Here and further the following notation is used

\[ G(z,\e )=e^{\textstyle \e\s(z,\bar{z})} a^{\textstyle -\e} \]
\newline
The first term in (\ref{10}) which is of zero order in the couplings expresses
the fact that even before the perturbation the original conformal theory has
some dependence on $\s$ through the conformal anomaly. In (\ref{10}) $c$ is
the central charge of the unperturbed conformal theory. The second term in
(\ref{10}), first order in the coupling, is the classical scale dependance
of the model resulting from its dimensionful couplings $g^{i}a^{-\e_{i}}$.
The third term in (\ref{10}) is just the second order iteration of this
classical dependence. The last term is an extra quantum mechanical scale
dependence arising from the presence of the cut off in the theory. Here
$\delta d(z,\om)$ is the variation in the geodesic distance of $z$ from $\om$
resulting from the variation of $\s$. To avoid any necessity to specify the
$\s$ dependence of the operator $A$, we can always chose the arbitrary
variation $\delta\s$ to vanish on the support of $A$.

  Up to first order in $z$ and also $\bar{z}$ derivatives express $d(z,\om)$
in terms of $\s$ near $\om$ as (see appendix)

\beq
 d(z,\om ) = e^{\s (\om)}|z-\om | (1+\frac{1}{2}(\pa\s (\om )
 (z-\om )+\bar{\pa}\s (\om )(\bar{z}-\bar{\om })) + \frac{1}{3}
 \pa\bar{\pa}\s (\om )|z-\om |^{2}~+ \label{11}
\eeq
\[ +\frac{1}{4}\pa\s (\om )\bar{\pa}\s (\om )|z-\om |^{2}) \]
\newline
which gives to this order for $\delta d(z,\om )$ of (\ref{10})

\beqa
 \delta d(z,\om)&=& e^{\s (\om)} \left\{ \left[ y\left( 1+\frac{1}{2}y
(e^{\th}\pa\s(\om)+ \right. \right. \right.
e^{-\th}\bar{\pa}\s(\om))+\frac{1}{3}y^{2}\pa\bar{\pa}\s(\om)+ \label{12} \\
&&+\frac{1}{4}y^{2}\pa\s(\om)\bar{\pa}\s(\om)\left)\left] \delta\s (\om)
+\frac{1}{2}y^{2}(e^{\th}\pa\delta\s(\om)+e^{-\th}\bar{\pa}\delta\s(\om))
\right.\right. \n
&&\left. +\frac{1}{4}y^{3}(\pa\delta\s\bar{\pa}\s+\pa\s\bar{\pa}\delta\s)
+\frac{1}{3}y^{3}\pa\bar{\pa}\delta\s(\om)\right\} \nonumber
\eeqa
\newline
where the real variables $y$ and $\theta$ stand for

\[ z-\om = ye^{\th} \]
{}~
The same change of variables can be made in the last term of (\ref{10}).
For any function of $y$ and $\theta$, $f(y,\theta )$,

\beq
\int d^{2}z \delta(d(z,\om)-a)f(y,\theta) = \int ydyd\theta~\delta(y-a)
\frac{\pa y(d,\theta)}{\pa d}\lfloor_{d=a} f(y,\theta) \label{13}
\eeq
\[ = \int d\theta~y(a,\theta)\frac{\pa y(a,\theta)}{\pa a}f(a,\theta) \]
{}~
Here $y(a,\theta)$ is the value of $|z-y|$ for which $d(z,\om)=a$ where the
phase of $z-\om$ is $\theta$. From (\ref{11}) we read

\beq
y(a,\theta)=e^{-\s(\om)}a\left( 1-\frac{1}{2}ae^{-\s(\om)}(\pa\s e^{\th}+
\bar{\pa}\s e^{-\th})-\frac{a^{2}}{3}e^{-2\s(\om)}\pa\bar{\pa}\s\right)
\label{14}
\eeq
\[ +~~higher~~ order~~ in~~ \pa\s,\bar{\pa}\s \]
{}~
and

\beq
\frac{\pa y(a,\theta)}{\pa a} = e^{-\s(\om)}[1-ae^{-\s(\om)}(\pa\s e^{\th}+
\bar{\pa}\s e^{-\th})-a^{2}e^{-2\s(\om)}\pa\bar{\pa}\s(\om)] \label{15}
\eeq
\newline
The last term of (\ref{10}) becomes , after applying (\ref{13}) with the
substitution of (\ref{12}),(\ref{14}) and (\ref{15})
\beqa
&&\frac{1}{2}\sum_{ij} g^{i}g^{j}\int d^{2}zd^{2}\om
G(z,\e_{i})G(\om,\e_{j})
<O_{i}(z,\bar{z})O_{j}(\om,\bar{\om})A>_{0}\times \label{16} \\
&&\times e^{-\s(\om)}a\left[ 1-\frac{1}{2}ae^{-\s}(\pa\s e^{\th}+
\bar{\pa}\s e^{-\th})-\frac{a^{2}}{3}e^{-2\s}\pa\bar{\pa}\s\right]
e^{-\s}\times \n
&&\times\left[ 1-ae^{-\s}(\pa\s e^{\th}+
\bar{\pa}\s e^{-\th})-a^{2}e^{-2\s}\pa\bar{\pa}\s\right]  \times \n
&&\times\left[ a\delta\s+\frac{1}{2}a^{2}e^{-\s(\om)}(e^{\th}\pa\delta\s+
e^{-\th}\bar{\pa}\delta\s)-\right.
\n
&&\left.
-\frac{1}{4}a^{3}e^{-2\s(\om)}(\pa\delta\s\bar{\pa}\s+\pa\s\bar{\pa}\delta\s)
+\frac{1}{3}a^{3}e^{-2\s(\om)}\pa\bar{\pa}\delta\s(\om)\right] \nonumber
\eeqa
{}~

   Expanding in (\ref{16}) $G(z,\e_{i})$ and the operator product around
$\om$ using (\ref{2}), the variation (\ref{10}) becomes
\beqa
&&\delta<A>=\frac{c}{6\pi}\int d^{2}\om~e^{2\s} R(\om,\bar{\om})
\delta\s(\om,\bar{\om})+ \label{17} \\
&&+\sum_{i}~\int d^{2}\om
G(\om,\e_{i})\e_{i}g^{i}<O_{i}(\om,\bar{\om})A>_{0}
\delta\s (\om,\bar{\om})+ \n
&&+\frac{1}{2}\sum_{ij}g^{i}g^{j}\int d^{2}zd^{2}\om
G(z,\e_{i})G(\om,\e_{j})(\e_{i}\delta\s (z,\bar{z})+\e_{j}
\delta\s (\om,\bar{\om}))\times \n
&&\times <O_{i}(z,\bar{z})O_{j}(\om,\bar{\om})A>_{0}\theta(d(z,\om )-a)+ \n
&&+\frac{1}{2}\sum_{ij}g^{i}g^{j}\int d^{2}\om
G(\om,\e_{i}+\e_{j})\int d\theta\sum_{k} C^{k}_{ij}\times \n
&&\times a^{\e_{i}+\e_{j}-\e_{k}}
\left[ 1+\e_{i}a(e^{\th}\pa\s(\om)+e^{-\th}\bar{\pa}\s(\om))+e_{i}a^{2}
\pa\bar{\pa}\s (\om)\right] e^{-\s (\om)}\times \n
&&\times\left[ 1-a(e^{\th}\pa\s(\om)+e^{-\th}\bar{\pa}
\s(\om))-a^{2}\pa\bar{\pa}\s (\om)\right]
\left[ \delta\s (\om)+\frac{1}{2}ae^{-\s(\om)}(e^{\th}\pa\delta\s(\om)+\right.
 \n
&&+e^{-\th}\bar{\pa}\delta\s(\om))-\frac{1}{4}a^{2}e^{-2\s(\om)}
(\pa\s(\om)\bar{\pa}\delta\s(\om)+\bar{\pa}\s(\om)\pa\delta\s(\om)) \n
&&\left.\left. +\frac{1}{3}a^{2}e^{-2\s(\om)}\pa\bar{\pa}\delta\s(\om)\right]
\times\left[ <O_{k}(\om)A>_{0}
+ae^{-\s(\om)}\right( 1-\frac{1}{2}ae^{-\s(\om)}(e^{\th}\pa\s(\om)+\right. \n
&&+\left.\left. e^{-\th}\bar{\pa}\s(\om))
-\frac{1}{3}a^{2}e^{-2\s(\om)}\pa\bar{\pa}\s(\om)\right)\right(
\ga_{ijk}e^{\th}\pa_{\om}<O_{k}(\om)A>_{0}+ \n
&&\left.\left. +\bar{\ga}_{ijk}e^{-\th}\pa_{\bar{\om}}<O_{k}(\om)A>_{0}\right)
+\ga_{ijk}\bar{\ga}_{ijk}a^{2}e^{-2\s}\pa_{\om}\pa_{\bar{\om}}
<O_{k}(\om)A>_{0}\right] \times \n
&&\times e^{-(\alpha_{ijk}+1)\s(\om)}\left[ 1-\frac{1}{2}(\alpha
_{ijk}+1)ae^{-\s(\om)}(e^{\th}\pa\s(\om)+e^{-\th}\bar{\pa}\s(\om))\right.  \n
&&\left. -\frac{1}{3}(\alpha_{ijk}+1)a^{2}e^{-2\s(\om)}\pa\bar{\pa}\s(\om)
\right]
e^{\th S_{k}} \nonumber
\eeqa
\newline
Here $\ga_{ijk}$ stands for $\ga^{(ijk)}_{1}$ of eq. (\ref{3}) namely

\beq
\ga_{ijk}=\frac{h_{i}-h_{j}+h_{k}}{2h_{k}} \label{18}
\eeq
\newline
and similarly for $\bar{\ga}_{ijk}$. Also $\alpha_{ijk}$ is a short notation
for

\beq
\alpha_{ijk} = h_{k}+\bar{h}_{k}-h_{i}-\bar{h}_{i}-h_{j}-\bar{h}_{j}
\label{19}
\eeq
\newline
and $S_{k}=h_{k}-\bar{h}_{k}$ is the spin of the operator $O_{k}$.
$\e_{k}=2-h_{k}-\bar{h}_{k}$ expresses the deviation of $O_{k}$ from
marginality.

   Eq. (\ref{17}) is valid up to first order in $\pa\s$,$\bar{\pa}\s$ and
$\pa\bar{\pa}\s$,and up to the first descendents of $O_{k}$ under left and
right hand side Virasoro algebra. In the same spirit we shall take into
account operators $O_{k}$ such that $S_{k}=0,+1,-1$. To this order integrating
by parts the $\pa\delta\s$ terms and performing the $\theta$ integration we
get

\beqa
&&\delta<A>=\frac{c}{6\pi}\int d^{2}\om~e^{2\s} R(\om,\bar{\om})
\delta\s (\om,\bar{\om})+ \label{20} \\
&&+\sum_{i}~\int d^{2}\om~
G(\om,\e_{i})\e_{i}g^{i}<O_{i}(\om,\bar{\om})A>_{0}
\delta\s (\om,\bar{\om})+ \n
&&+\frac{1}{2}\sum_{ij}g^{i}g^{j}\int d^{2}zd^{2}\om
G(z,\e_{i})G(\om,\e_{j})\times \n
&&\times
(\e_{i}\delta\s (z,\bar{z})+\e_{j}\delta\s (\om,\bar{\om}))
<O_{i}(z,\bar{z})O_{j}(\om,\bar{\om})A>_{0}\theta(d(z,\om )-a)+ \n
&&+\left. \left. \pi\sum_{ijk}C^{k}_{ij}\int d^{2}\om g^{i}g^{j}
\right\{ \right[ G(\om,\e_{k})<O_{k}(\om)A>_{0}+ \n
&&+(\ga^{2}_{ijk}-\ga_{ijk}+\frac{1}{3})G(\om,\e_{k}-2)
\pa_{\om}{\pa}_{\bar{\om}}<O_{k}(\om)A>_{0}- \n
&&-2h_{k}(\ga^{2}_{ijk}-\ga_{ijk}+\frac{1}{3})G(\om,\e_{k}-2)
(\pa\s(\om)\bar{\pa}<O_{k}(\om)A>_{0}+\bar{\pa}\s(\om)\pa<O_{k}(\om)A>_{0})+ \n
 && \left. +\frac{1}{6}(\e_{i}+\e_{j}+\e_{k}-2)G(\om,\e_{k}-2)
\pa\bar{\pa}\s(\om)<O_{k}(\om)A>_{0}\right] \delta_{S_{k},0}+ \n
&& +G(\om,\e_{k}-1)\left[ \frac{\bar{h}_{i}-\bar{h}_{j}}
{2\bar{h}_{k}}\bar{\pa}<O_{k}(\om)A>_{0}-(\bar{h}_{i}-\bar{h}_{j})\bar{\pa}\s
<O_{k}(\om)A>_{0}\right] \delta_{S_{k},1}+ \n
&&\left. +G(\om,\e_{k}-1)\left[ \frac{h_{i}-h_{j}}
{2h_{k}}\pa<O_{k}(\om)A>_{0}-(h_{i}-h_{j})\pa\s
<O_{k}(\om)A>_{0}\right] \delta_{S_{k},-1}\right\} \delta\s(\om) \nonumber
\eeqa
\newline
In the gauge (\ref{5}), due to (\ref{6}), the covariant derivative of $O_{k}$
is

\beq
DO_{k}=\pa O_{k}-2h_{k}\pa\s O_{k} \label{21}
\eeq
\newline
and, to first order in $\s$ derivatives,

\beq
D\bar{D}O_{k}=\pa\bar{\pa}O_{k}-2h_{k}\pa\s\bar{\pa}O_{k}
-2\bar{h}_{k}\bar{\pa}\s\pa O_{k}-2\bar{h}_{k}\pa\bar{\pa}\s O_{k} \label{22}
\eeq
\newline
   Note that the coefficients in (20) of $\pa <O_{k}A>$ and $\pa\s
<O_{k}A>$ and also of $\pa\bar{\pa}<O_{k}A>$ and $\pa\s\bar{\pa} <O_{k}A>$ are
related precisely such that their combination are expressible in terms of
covariant derivatives. Also the term involving $\pa\bar{\pa}\s$, is
expressible in terms of the scalar world sheet curvature due to (\ref{7}).
This results of course from our coordinate independent regularization
procedure.

   The variation $\delta\s$ induces a change in the local cut off scale
$ae^{-\s}$. In the spirit of the renormalization group, one looks for a
corresponding local change in the coupling constants $g^{i}\rightarrow g^{i}
+\beta^{i}\delta\s$ which will compensate for the change of scale such that
$<A>$ calculated with the new couplings for the new cut off scale will equal to
$<A>$ with the old couplings for the original scale. The appearance in $\delta
<A>$ of (20) of terms involving $D<O_{k}A>,~D\bar{D}<O_{k}A>$ and $R
<O_{k}A>$ implies that the renormalization should not only change the values
of existing couplings to primary operators but also induce new types of
couplings to descendent operators of the type $DO_{k}$ and $D\bar{D}O_{k}$
and to dilatonic type operators of the form $RO_{k}$. Let us denote the
coupling to the operator $DO_{k}$ by $g^{k}_{z}$, that of $\bar{D}O_{k}$ by
$g^{k}_{\bar{z}}$, that of $D\bar{D}O_{k}$ by $g^{k}_{z\bar{z}}$, and the
dilaton coupling to $RO_{k}$ by $\tilde{g}^{k}$.The requirement of the local
renormalization flow to preserve physical quantity under change of scale is
expressed by the corresponding Callan Symanzik equation

\beq
\delta <A>+\int d^{2}\om\sum_{k}\left(\beta^{k}\frac{\delta}{\delta g^{k}}+
\beta^{k}_{z}\frac{\delta}{\delta g^{k}_{z}}+\beta^{k}_{\bar{z}}
\frac{\delta}{\delta g^{k}_{\bar{z}}}+\right. \label{23}
\eeq
\[ \left. +\beta^{k}_{z\bar{z}}\frac{\delta}{\delta g^{k}_{z\bar{z}}}+
\tilde{\beta}^{k}\frac{\delta}{\delta \tilde{g}^{k}}\right) <A>\delta\s
(\om)~=~0 \]
\newline
Since $\frac{\delta <A>}{\delta g^{k}}=-a^{-\e_{k}}e^{\e_{k}\s(\om)}<O_{k}(\om)
A>$, eq. (\ref{23}) reads

\beqa
&&\delta <A>=\int d^{2}\om \sum_{k}\left\{ \beta^{k}G(\om,\e_{k})
<O_{k}(\om)A>+ \right.
\label{24} \\
&&+G(\om,\e_{k}-1)(\beta^{k}_{z}D_{\om}<O_{k}(\om)A>+
 \beta^{k}_{\bar{z}}\bar{D}_{\om}<O_{k}(\om)A>)+ \n
&&+\beta^{k}_{z\bar{z}}\times G(\om,\e_{k}-2)
\left.
D_{\om}\bar{D}_{\om}<O_{k}(\om)A>-
\tilde{\beta}^{k}G(\om,\e_{k}-2)\pa\bar{\pa}\s(\om)
<O_{k}(\om)A>\right\} \delta\s(\om) \nonumber
\eeqa

   We have already an expression for $\delta<A>$ up to second order in the
perturbation in eq.(\ref{20}). Expand also each $\beta$-function and each
expectation value in the r.h.s. of (\ref{24}) to this order and equate term
by term to (\ref{20}) to get,

\beqa
&\beta^{k}&=\e_{k}g^{k}+\delta_{S_{k},0}\pi\sum_{ij} C^{k}_{ij}g^{i}g^{j}
 \label{25} \\
&\beta^{k}_{z}&=\delta_{S_{k},-1}\pi\sum_{ij} C^{k}_{ij}g^{i}g^{j}\frac{
h_{i}-h_{j}}{2h_{k}} \n
&\beta^{k}_{\bar{z}}&=\delta_{S_{k},1}\pi\sum_{ij} C^{k}_{ij}g^{i}g^{j}\frac{
\bar{h}_{i}-\bar{h}_{j}}{2\bar{h}_{k}} \n
&\beta^{k}_{z\bar{z}}&=\delta_{S_{k},0}\pi\sum_{ij} C^{k}_{ij}g^{i}g^{j}
(\ga^{2}_{ijk}-\ga_{ijk}+\frac{1}{3}) \n
&\tilde{\beta}^{k}&=\frac{c}{6\pi}\delta_{k,0}
-\delta_{S_{k},0}\pi\sum_{ij} C^{k}_{ij}g^{i}
g^{j}\left[ \frac{1}{6}(\e_{i}+\e_{j}+\e_{k}-2)+2h_{k}
(\ga^{2}_{ijk}-\ga_{ijk}+\frac{1}{3}) \right] \nonumber
\eeqa
\newline
where $O_{0}$ stands for the unit operator.

   The first equation in (\ref{25}) is the renormalization of primary
couplings found in \cite{5,6}, the other equations are its generalization to
leading descendent and dilatonic operators. We see from our analysis, that
even if originally the perturbation is primary, descendent operators, in
principle of any order, and couplings to any power of the world sheet curvature
 are induced by the renormalization group. Induced derivative type operators
like $DO_{k}$ or $D\bar{D}O_{k}$ are certainly significant in the local
renormalization framework where they have non constant coupling. In our way
of regularization they have non trivial influence beyond first order even for
a background with constant couplings in spite of their appearance as total
derivatives. This is because of the boundary terms at the edge of the small
disc of radius $a$ present in this scheme.

{}~

\section{Inclusion of non primary couplings}

   Since non primary and dilatonic couplings are induced in the process of
renormalization flow, in order to follow this flow one has to know the $\beta$-
function for initial perturbations of this type. Here we will repeat the
calculation of the previous section for a perturbation more general than that
of eq.(\ref{8}). We take the perturbed action to be

\beqa
&&S=S_{0}+\int d^{2}z\sum_{i}\left(
e^{\e_{i}\s (z)}a^{-\e_{i}}g^{i} O_{i}(z,\bar{z})+e^{(\e_{i}-1)\s (z)}
a^{1-\e_{i}}(g^{i}_{z}D_{z}O_{i}(z,\bar{z})+\right. \label{26} \\
&&\left.+g^{i}_{\bar{z}}D_{\bar{z}}O_{i}(z,\bar{z}))
 +g^{i}_{z\bar{z}}e^{(\e_{i}-2)\s (z)}a^{2-\e_{i}}
D_{z}\bar{D}_{\bar{z}}O_{i}(z,\bar{z})+
\tilde{g}^{i}e^{\e_{i}\s(z)}a^{2-\e_{i}}R(z,\bar{z})O_{i}(z,\bar{z})
 \right) \nonumber
\eeqa
\newline
We assume constant couplings, for general covariance of the action they should
couple to scalars. Hence $g^{i}_{z}$ is non zero only for an operator $O_{i}$
of spin -1, $g^{i}_{\bar{z}}$ exists only for $O_{i}$ of spin 1 and all the
remaining couplings in (\ref{26}) involve operators of spin 0.

   Up to second order in the couplings and first order in $z$ and $\bar{z}$
derivatives, the expectation of an arbitrary operator $A$ is

\beqa
&&<A>=<A>_{0}+\int d^{2}\om\sum_{i}\left[
G(\om,\e_{i})g^{i}<O_{i}(\om)A>_{0}+\right. \label{27} \\
&&+G(\om,\e_{i}-1)(g^{i}_{z}D_{\om}<O_{i}(\om)A>_{0}+
g^{i}_{\bar{z}}\bar{D}_{\bar{\om}}<O_{i}(\om)A>_{0})+ \n
&&+G(\om,\e_{i}-2)(g^{i}_{z\bar{z}}
\left. D_{\om}\bar{D}_{\bar{\om}}<O_{i}(\om)A>_{0}-
\tilde{g}^{i}\pa\bar{\pa}\s(\om)<O_{i}(\om)A>_{0}) \right]+ \n
&&+\sum_{ij}\int d^{2}zd^{2}\om ~\theta(d(z,\om)-a)
 \left\{ \frac{1}{2}G(z,\e_{i})G(\om,\e_{j})g^{i}g^{j}
<O_{i}(z)O_{j}(\om)A>_{0}+\right.\n
&&+G(z,\e_{i})G(\om,\e_{j}-1)g^{i}(g^{j}_{z}D_{\om}+
g^{j}_{\bar{z}}\bar{D}_{\bar{\om}})<O_{i}(z)O_{j}(\om)A>_{0}+ \n
&&+G(z,\e_{i})G(\om,\e_{j}-2)
\left[ g^{i}g^{j}_{z\bar{z}}D_{\om}\bar{D}_{\bar{\om}}
 <O_{i}(z)O_{j}(\om)A>_{0}- \right. \n
&&\left. -g^{i}\tilde{g}^{j}\pa\bar{\pa}\s(\om)
<O_{i}(z)O_{j}(\om)A>_{0}\right] + \n
&& \left. +g^{i}_{z}g^{j}_{\bar{z}}G(z,\e_{i}-1)G(\om,\e_{j}-1)
D_{z}\bar{D}_{\bar{\om}}<O_{i}(z)O_{j}(\om)A>_{0} \right\} \nonumber
\eeqa
\newline
Again, vary $\s$ into $\s+\delta\s$ with arbitrary $\delta\s$ which vanishes
on the support of the operator $A$, find $\delta<A>$ to second order and
substitute into the Callan Symanzik equation (\ref{23}) to determine the
$\beta$
-functions perturbatively. To first order one has the dimension counting
result

\beqa
\beta^{k}&=&\e_{k}g^{k} \label{28} \\
\tilde{\beta}^{k}&=&\frac{c}{6\pi}\delta_{k,0}+(\e_{k}-2)\tilde{g}^{k} \n
\beta^{k}_{z}&=&\beta^{k}_{\bar{z}}~=~0 \n
\beta^{k}_{z\bar{z}}&=&-\tilde{g}^{k} \nonumber
\eeqa
\newline
Note that the explicit $\s$ dependence in the coefficient $e^{(\e_{i}-1)\s}$
of $g^{i}_{z}DO_{i}$ exactly cancels against the extra $\s$ dependence in
the covariant derivative, eq. (\ref{21}), to give no $g^{k}_{z}$ dependence
in $\beta$ to first order , as should be expected since to this order the
coupling $DO_{k}$ is indeed a spurious total derivative. The explicit $\s$
dependence in the $R$ term contributes to the first order $\beta$-function
of $g^{k}_{z\bar{z}}$.

   The presence of the dilatonic and descendent couplings in (\ref{26})
introduces additional sources of $\s$ dependence compared to that of the
previous section. For example let us consider the second order contribution
to $\delta<A>$ proportional to $g^{i}\tilde{g}^{j}$. Varying this term in
(\ref{27}) we get

\beqa
&&\delta<A>=-\sum_{ij}g^{i}\tilde{g}^{j}\int d^{2}zd^{2}\om
G(z,\e_{i})G(\om,\e_{j}-2)\left[ (\e_{i}\delta\s(z)+
(\e_{j}-2)\delta\s(\om))\times  \right. \label{29} \\
&&\times \pa\bar{\pa}\s(\om)<O_{i}(z)O_{j}(\om)A>_{0}\theta(d(z,\om)-a)+
\pa\bar{\pa}\s(\om)<O_{i}(z)O_{j}(\om)A>_{0}\times \n
&&\left.\times\delta(d(z,\om)-a)\delta d(z,\om)+
 \pa\bar{\pa}\delta\s(\om)<O_{i}(z)O_{j}(\om)A>_{0}\theta(d(z,\om)-a)
\right]+ \n
&&+~other~terms~not~proportional~to~g^{i}\tilde{g}^{j} \nonumber
\eeqa
\newline
The first two terms are familiar from previous section. The first term of
$\delta<A>$ in (\ref{29}) is taken care of by the first order piece of the
$\beta$-functions $\beta^{i}$ and $\tilde{\beta}^{j}$ of eq. (\ref{28}) when
substituted into the Callan Symanzik equation (\ref{24}). The second term
which involves $\delta d(z,\om)$ is treated exactly as in previous section
and we can copy from eq. (\ref{20}) after integrating over the $\delta$-
function and utilizing operator product expansion the form for this term,

\beq
-2\pi\sum_{ijk}C^{k}_{ij}\int d^{2}\om~g^{i}\tilde{g}^{j}
G(\om,\e_{k}-2)\pa\bar{\pa}\s(\om)<O_{k}(\om)A>_{0} \label{30}
\eeq
\newline
thus contributing to $\tilde{\beta}^{k}$.

   The third term in (\ref{29}) comes from the explicit dependence of the
dilatonic coupling on $\s$ through the curvature factor. Integrating
$\pa_{\om}\bar{\pa}_{\bar{\om}}\delta\s$ by parts this term gives rise to

\beqa
&&-\sum_{ij}g^{i}\tilde{g}^{j}\int d^{2}zd^{2}\om~
G(z,\e_{i})G(\om,\e_{j}-2)\times \label{31} \\
&&\times \left[ D_{\om}\bar{D}_{\bar{\om}}<O_{i}(z)O_{j}(\om)A>_{0}\right.
\theta(d(z,\om)-a)+D_{\om}<O_{i}(z)O_{j}(\om)A>_{0}\bar{\pa}_{\bar{\om}}
\theta(d(z,\om)-a)+ \n
&&+\bar{D}_{\bar{\om}}<O_{i}(z)O_{j}(\om)A>_{0}\pa_{\om}\left.
\theta(d(z,\om)-a) +<O_{i}(z)O_{j}(\om)A>_{0}\pa_{\om}\bar{\pa}_{\bar{\om}}
\theta(d(z,\om)-a)\right] \nonumber
\eeqa
\newline
The first term in (\ref{31}), which involves double integration over $z$ and
$\om$ with no $\delta$-function, does not induce a new second order $\beta$-
function. Rather, it is taken care of by the first order contribution to
$\beta^{j}_{z\bar{z}}$ in (\ref{28}), multiplied by the first order correction
to $D\bar{D}<O_{j}A>$ in the term $\beta^{j}_{z\bar{z}}D\bar{D}<O_{j}A>$ of
the Callan Symanzik equation (\ref{24}). For the remaining 3 terms in
(\ref{31}) use the identities

\beqa
&&\pa_{\om}\theta(d(z,\om)-a)=\delta(d(z,\om)-a)\pa_{\om}d(z,\om) \label{32} \\
&&\pa_{\om}\bar{\pa}_{\bar{\om}}\theta(d(z,\om)-a)=\delta'(d(z,\om)-a)
\pa_{\om}d(z,\om)\bar{\pa}_{\bar{\om}}d(z,\om)+\delta(d(z,\om)-a)
\pa_{\om}\bar{\pa}_{\bar{\om}}d(z,\om) \nonumber
\eeqa
\newline
The $\delta$-functions on the r.h.s. of (\ref{32}) make sure that in the last
3 terms in (\ref{31}) the double integral on $z$ and $\om$ is effectively only
integrals over $\om$ which can be canceled by local second order counter terms
in $\beta$-functions. To compute the required terms pass again to the variables
 $ye^{\th}=z-\om$ and integrate the $\delta$-function using (\ref{13}) and the
formula

\beq
\int dy~f(y,\theta)\delta'(d(y,\theta)-a)=-
\frac{\pa^{2}y(a,\theta)}{\pa a^{2}} f(y(a,\theta))-\left(
\frac{\pa y(a,\theta)}{\pa a}\right)^{2}f'(y(a,\theta)) \label{33}
\eeq
\newline
valid for any function of $y$ and $\theta$. For the quantities
$\pa_{\om}d(z,\om)$ and $\pa_{\om}\bar{\pa}_{\bar{\om}}d(z,\om)$ appearing in
(\ref{32}), derive from (\ref{11}) their expansion in $\s$ derivatives

\beqa
&&\pa_{\om} d(z,\om ) =-\frac{1}{2}e^{\s
(\om)}e^{-\th}[1-\frac{y}{2}(\pa\s(\om)
e^{\th}-\bar{\pa}\s(\om)e^{-\th})] \label{34} \\
&&\pa_{\om}\bar{\pa}_{\bar{\om}}d(z,\om )=\frac{e^{\s(\om)}}{4y}[1-\frac{y}{2}
(\pa\s(\om)e^{\th}+\bar{\pa}\s(\om)e^{-\th})+y^{2}\pa\bar{\pa}\s(\om)]
\nonumber
\eeqa

   As an example consider the second term in (\ref{31}) involving
$D_{\om}<O_{i}(z)O_{j}(\om)A>_{0}$. We have

\beq
-\sum_{ij}\int d^{2}zd^{2}\om G(z,\e_{i})G(\om,\e_{j}-2)D_{\om}
<O_{i}(z)O_{j}(\om)A>_{0}\bar{\pa}_{\om}\theta(d(z,\om)-a)\delta\s(\om)
 \label{35}
\eeq
\newline
In terms of $y$ and $\theta$

\[ \pa_{\om}=-\frac{e^{-\th}}{2}\frac{\pa}{\pa y}-\frac{1}{2y}\frac{\pa}
{\pa e^{\th}} \]
\newline
Apply this to the operator product formula,

\beqa
&& <O_{i}(z)O_{j}(\om)A>_{0}=\sum_{k}C^{k}_{ij}y^{\alpha_{ijk}}
e^{\th S_{ijk}}\left[ <O_{k}(\om)A>_{0}+ \right.\n
&&+y(\ga_{ijk}e^{\th}\pa_{\om}<O_{k}(\om)A>_{0}+
\bar{\ga}_{ijk}e^{-\th}\pa_{\bar{\om}}<O_{k}(\om)A>_{0})
\left. +y^{2}\ga_{ijk}\bar{\ga}_{ijk}\pa_{\om}\bar{\pa}_{\bar{\om}}
<O_{k}(\om)A>_{0}\right] \nonumber
\eeqa
\newline
with $S_{ijk}=S_{k}-S_{i}-S_{j}$ to get

\beqa
&&D_{\om}<O_{i}(z)O_{j}(\om)A>_{0}=(\pa_{\om}-2h_{j}\pa\s(\om))
<O_{i}(z)O_{j}(\om)A>_{0}= \label{36} \\
&&\left. =-\sum_{k}C^{k}_{ij}\right\{ \left[\frac{1}{2}(\alpha_{ijk}+S_{ijk})+
2h_{j}y e^{\th}\pa\s(\om)\right] <O_{k}(\om)A>_{0}+ \n
&&+\left[ \left( \frac{1}{2}(\alpha_{ijk}+S_{ijk}+2)+
2h_{j}y e^{\th}\pa\s(\om)\right) \ga_{ijk}-1\right] y e^{\th}
\pa_{\om}<O_{k}(\om)A>_{0}+ \n
&&+ \left( \frac{1}{2}(\alpha_{ijk}+S_{ijk})+
2h_{j}y e^{\th}\pa\s(\om)\right) \bar{\ga}_{ijk}y e^{-\th}
\bar{\pa}_{\bar{\om}}<O_{k}(\om)A>_{0}+ \n
&& +y^{2}\left[ \left( \frac{1}{2}(\alpha_{ijk}+S_{ijk}+2)+
2h_{j}y e^{\th}\pa\s(\om)\right) \ga_{ijk}\bar{\ga}_{ijk}-(\ga_{ijk}e^{2\th}+
\bar{\ga}_{ijk})\right]\times \n
&& \times \pa_{\om}\bar{\pa}<O_{k}(\om)A>_{0}\left\}
y^{\alpha_{ijk}
-1}e^{\th(S_{ijk}-1)}\right. \nonumber
\eeqa
\newline
We have then for (\ref{35})

\beqa
&&-\sum_{ij}g^{i}\tilde{g}^{j}\int d^{2}zd^{2}\om G(\om,\e_{i}+\e_{j}-2)
\int d\theta
\left[ 1+\e_{i}y(a,\theta)(\pa\s(\om)e^{\th}+\bar{\pa}\s(\om)e^{-\th})+\right.
\label{37} \\
&& +\e_{i}y^{2}(a,\theta)\pa\bar{\pa}\s(\om)\left] e^{\th}\right. \left(
-\frac{e^{\s(\om)}}{2}\right)
\left[ 1+\frac{y(a,\theta)}{2}(\pa\s(\om)e^{\th}-\bar{\pa}\s(\om)e^{-\th})
\right]\times \n
&& \times e^{-\s(\om)}\left[ 1-e^{-\s(\om)}(\pa\s e^{\th}+\bar{\pa}\s e^{-\th})
-e^{-2\s(\om)}\pa\bar{\pa}\s \right](-C^{k}_{ij})y^{\alpha_{ijk}}(a,\theta)
e^{\th(S_{ijk}-1)}\times \n
&&\times \left[ \left(\frac{1}{2}(\alpha_{ijk}+S_{ijk})-2h_{j}y(a,\theta)
e^{\th}
\pa\s\right) <O_{k}(\om)A>_{0}+\right. \n
&&+\left(\left(\frac{1}{2}(\alpha_{ijk}+S_{ijk}+2)+2h_{j}y(a,\theta)e^{\th}
\pa\s\right)\ga_{ijk}-1\right) y(a,\theta)e^{\th}\pa_{\om}<O_{k}(\om)A>_{0}+
\n
&&+\frac{1}{2}\left( (\alpha_{ijk}+S_{ijk})+2h_{j}y(a,\theta)e^{\th}
\pa\s\right)\bar{\ga}_{ijk} y(a,\theta)e^{-\th}\bar{\pa}_{\bar{\om}}<O_{k}(\om)
A>_{0}+\n
&&+\left(\left(\frac{1}{2}(\alpha_{ijk}+S_{ijk}+2)+2h_{j}y(a,\theta)e^{\th}
\pa\s\right)\ga_{ijk}\bar{\ga}_{ijk}-\ga_{ijk}e^{2\th}-\bar{\ga}_{ijk}\right)
\times \n
&&\left. \times y^{2}(a,\theta)\pa_{\om}\bar{\pa}_{\bar{\om}}
<O_{k}(\om)A>_{0}\right] \nonumber
\eeqa
\newline
The first factor in (\ref{37}) is the expansion of $e^{\e_{i}\s(z)}$ about
$\om$, the second factor is $\bar{\pa}d(z,\om)$ eq. (\ref{34}). The third
factor
 is $\frac{\pa y(a,\theta)}{\pa a}$ from eq. (\ref{15}), the Jacobian resulting
 from integrating $\delta(d(y,\theta)-a)$ over $y$. The last factor is $D_{\om}
<O_{i}(z)O_{j}(\om)A>_{0}$ from (\ref{36}) multiplied by an extra $y$ from the
measure $yd\theta$ of polar integration. Due to the $\delta(d(y,\theta)-a)$
for every $y$ of (\ref{37}) we have to substitute $y(a,\theta)$ of eq.
(\ref{14}), the value of the coordinate $y$ for which $d(y,\theta)=a$.
Substituting (\ref{14}) and performing the $\theta$ integration, keeping
operators of spin 0,+1,-1, and first order $\s$ derivatives, we get for
(\ref{37}),

\beqa
&&-(2\pi)\sum_{ijk}C^{k}_{ij}g^{i}\tilde{g}^{j}\int d^{2}\om \left\{
\left[ \frac{\alpha
_{ijk}}{4}G(\om,\e_{k})<O_{k}(\om)A>_{0}+\right. \right. \label{38} \\
&&+\frac{1}{4}\left( (\alpha_{ijk}+2)\ga_{ijk}-2\right)\bar{\ga}_{ijk}
G(\om,\e_{k}-2)\pa\bar{\pa}<O_{k}(\om)A>_{0}+ \n
&&+\frac{1}{4}\left( (\alpha_{ijk}+2)\ga_{ijk}-2 \right) \left( \e_{i}-\frac{
\alpha_{ijk}+4}{2}\right) G(\om,\e_{k}-2)(\pa\s\bar{\pa}<O_{k}(\om)A>_{0}+
\bar{\pa}\s\pa<O_{k}(\om)A>_{0})+ \n
&&\left. +\frac{1}{4}\alpha_{ijk}\left( \e_{i}-1-\frac{\alpha_{ijk}}{3}\right)
 G(\om,\e_{k}-2)\pa\bar{\pa}\s(\om)<O_{k}(\om)A>_{0}\right]
\delta_{S_{ijk},0}+ \n
&&+\left[ \frac{1}{4}\left( (\alpha_{ijk}+1)\ga_{ijk}-2\right)G(\om,\e_{k}-1)
\pa<O_{k}(\om)A>_{0}+\right.  \n
&&+\left. \frac{1}{4}\left( (\alpha_{ijk}-1)\left( \e_{i}-\frac{\alpha_{ijk}+1}
{2}
\right)-4h_{j}\right)G(\om,\e_{k}-1)\pa\s <O_{k}(\om)A>_{0}\right]\delta_
{S_{ijk},-1}+ \n
&&+\left[ \frac{1}{4}(\alpha_{ijk}+1)\bar{\ga}_{ijk}G(\om,\e_{k}-1)\bar{\pa}
<O_{k}(\om)A>_{0}+\right.  \n
&&+\left. \left. \frac{1}{4}(\alpha_{ijk}+1)\left( \e_{i}-\frac{\alpha_{ijk}+3}
{2}
\right)G(\om,\e_{k}-1)\bar{\pa}\s <O_{k}(\om)A>_{0}\right]\delta_
{S_{ijk},1}\right\}\delta\s(\om) \nonumber
\eeqa
\newline
Again, as they should the coefficients of $\pa <O_{k}A>$ and of $\pa\s
<O_{k}A>$
 are related by (\ref{21}) and (\ref{22}) to form covariant world sheet
derivatives and (\ref{38}) becomes:

\beqa
&&-(2\pi)\sum_{ijk}C^{k}_{ij}g^{i}\tilde{g}^{j}\int d^{2}\om \left\{ \left[
\frac{\alpha
_{ijk}}{4}G(\om,\e_{k})<O_{k}(\om)A>_{0}+\right. \right. \label{39} \\
&&+\frac{1}{4}\left( (\alpha_{ijk}+2)\ga_{ijk}-2\right)\bar{\ga}_{ijk}
G(\om,\e_{k}-2)D\bar{D}<O_{k}(\om)A>_{0}+ \n
&&+\frac{1}{4}\left(\alpha_{ijk}\left(\e_{i}-\frac{\alpha_{ijk}+3}{3}\right) -
\left( (\alpha_{ijk}+2)\ga_{ijk}-2 \right)\times\right. \n
&&\times\left.\left.\left( \e_{i}-\frac{\alpha_{ijk}+4}{2}\right)\right)
 G(\om,\e_{k}-2)(\pa\bar{\pa}\s <O_{k}(\om)A>_{0}\right]\delta_{S_{ijk},0}+\n
&&+ \frac{1}{4}\left( (\alpha_{ijk}+1)\ga_{ijk}-2\right)G(\om,\e_{k}-1)
D<O_{k}(\om)A>_{0}\delta_{S_{ijk},-1}+ \n
&&+\left. \frac{1}{4}(\alpha_{ijk}+1)\bar{\ga}_{ijk}G(\om,\e_{k}-1)\bar{D}
<O_{k}(\om)A>_{0}\delta_{S_{ijk},1}\right\}\delta\s(\om) \nonumber
\eeqa
\newline
{}From (\ref{39}) the contribution of this particular term in (\ref{31}) to the
various second order $\beta$-functions are explicitly read off. The remaining
terms in (\ref{31}) are treated in the same way.

   In a similar fashion the terms in $<A>$ of eq. (\ref{27}) involving
$g^{i}g^{j}_{z}$, $g^{i}_{z}g^{j}_{\bar{z}}$ or $g^{i}g^{j}_{z\bar{z}}$ have
also extra contributions to $\delta <A>$ besides that coming from $\delta
d(z,\om)$ discussed in previous section. All these terms contain $\s$
derivatives inside the covariant derivatives, which contribute derivatives of
$\delta\s$ after the variation. Exactly as in eq. (\ref{31}) when these
$\delta\s$ derivatives are integrated by parts, one gets boundary terms from
the edges of integration region at the cut off circle of radius $a$. These
boundary terms are treated in the same way as we did to eq. (\ref{31}) to
extract their contributions to $\beta$-functions.

   Collecting all contributions we have our final results for the $\beta$-
functions induced by a perturbation of a general conformal theory by primary,
first order descendent, and first order dilatonic operators with uniform
coupling constants,

\beqa
\beta^{k}&=&\e_{k}g^{k}+2\pi\sum_{ij}C^{k}_{ij}\left[\frac{1}{2}g^{i}g^{j}+
\frac{\alpha}{4}g^{i}\tilde{g}^{j}+\frac{1}{4}\alpha(\alpha-\e_{j}+2)
g^{i}g^{j}_{z\bar{z}}+\right. \label{40} \\
&&\left. +\frac{1}{2}(\e_{j}-\alpha-2)g^{i}(g^{j}_{z}+g^{j}_{\bar{z}})+
\frac{1}{4}((\e_{i}+\e_{j}-2)-\alpha)\alpha g^{i}_{z}g^{j}_{\bar{z}}\right] ,
 \n
\beta^{k}_{z\bar{z}}&=&\tilde{g}^{k}+2\pi\sum_{ij}C^{k}_{ij}\left\{\frac{1}{2}
\left[\ga^{2}-\ga+\frac{1}{3}\right] g^{i}g^{j}+\left[\frac{\ga^{2}(\alpha+2)}
{4}-\ga\right]g^{i}\tilde{g}^{j}+\right. \n
&&+\left[\left( 1-\frac{\ga (\alpha+2)}{2}\right)(1-\ga)\frac{\alpha+2}{2}-
(\e_{j}-2)\ga^{2}\frac{\alpha+2}{4}+\frac{\alpha^{2}}{12}\right] g^{i}g^{j}_
{z\bar{z}}+ \n
&&+\left[\left( 1-\frac{\ga (\alpha+3)}{2}\right)(\bar{\ga}-\frac{1}{2})
+\left(\frac{\bar{\ga}}{2}-\frac{1}{3}\right)
\frac{\alpha+1}{2}+\frac{\e_{j}-1}{2}\ga\bar{\ga}\right]g^{i}g^{j}_
{z}+ \n
&&+\left[\left( 1-\frac{\bar{\ga}(\alpha+3)}{2}\right)(\ga-\frac{1}{2})
+\left(\frac{\ga}{2}-\frac{1}{3}\right)
\frac{\alpha+1}{2}+\frac{\e_{j}-1}{2}\ga\bar{\ga}\right]g^{i}g^{j}_
{\bar{z}}+ \n
&&+\left[\left( 1-\frac{\bar{\ga}(\alpha+2)}{2}\right)\left(\frac{\ga(\alpha+
2)}{2}-\frac{\alpha}{4}\right)+\frac{\ga\alpha(\alpha+2)}{8}-\frac{\alpha^{2}}
{12}+\right. \n
&&+\left.\left. (\e_{j}-1)\ga\bar{\ga}\frac{\alpha+2}{4}+(\e_{i}-1)(1-\ga)(1-
\bar{\ga})\frac{\alpha+2}{4}\right] g^{i}_{z}g^{j}_{\bar{z}}\right\} ,
\n
\tilde{\beta}^{k}&=&\frac{c}{6\pi}\delta_{k,0}+
(\e_{k}-2)\tilde{g}^{k}+2\pi\sum_{ij}C^{k}_{ij}\left\{
\left[\e_{i}-1-\frac{\alpha+1}{3}-\ga\left(\e_{i}-1-\frac{\alpha+2}{2}\right)
\right]\frac{g^{i}g^{j}}{2}+\right. \n
&&+\left[ 1+\frac{\alpha+2}{2}\left(\e_{i}\frac{1-\ga}{2}+\ga\frac{\alpha+4}
{4}-1\right)-\frac{\alpha^{2}}{12}\right] g^{i}\tilde{g}^{j}+ \n
&&+\left[\frac{\alpha+2}{2}\left( (\e_{j}-2)+\frac{\alpha^{2}}{12}-(\e_{k}-2)
(1-\ga)\left( 1-\frac{\ga(\alpha+2)}{2}\right)\right)+\frac{\e_{j}-2}{4}\left(
\e_{i}-\frac{\alpha}{3}+2-\right.\right. \n
&&-\left.\left. \e_{i}(\alpha+3)+\frac{(\alpha+1)(\alpha+6)}{3}+\left(\e_{i}-
\frac{\alpha+4}{2}\right)\ga(\alpha+2)\right)\right] g^{i}g^{j}_{z\bar{z}}+ \n
&&+\left[-\frac{\alpha+1}{2}\left(\e_{i}-\frac{\alpha+3}{3}\right)-\left( 1-
\frac{\ga(\alpha+3)}{2}\right)\left(\e_{i}-\frac{\alpha+3}{2}\right)
+\right. \n
&&+\left.\frac{\e_{j}-1}{2}\left(\e_{i}-\frac{\alpha+4}{3}
-\bar{\ga}\left(\e_{i}
-\frac{\alpha+4}{2}\right)\right)\right] g^{i}g^{j}_{z}+ \n
&&+\left[-\frac{\alpha+1}{2}\left(\e_{i}-\frac{\alpha+3}{3}\right)-\left( 1-
\frac{\bar{\ga}(\alpha+3)}{2}\right)\left(\e_{i}-\frac{\alpha+3}{2}\right)
+\right. \n
&&+\left.\frac{\e_{j}-1}{2}\left(\e_{i}-\frac{\alpha+4}{3}-\ga\left(\e_{i}
-\frac{\alpha+4}{2}\right)\right)\right] g^{i}g^{j}_{\bar{z}}+ \n
&&+\left[\left( 1-\frac{\alpha}{2}\right)(\e_{i}-1)-\frac{\alpha+2}{3}-\left(
1-\bar{\ga}\frac{\alpha+2}{2}\right)\left(\e_{i}-1
+\frac{\alpha}{2}\left(\e_{i}-
\frac{\alpha+4}{2}\right)\right)+\right. \n
&&+(\e_{j}-1)(\e_{i}-1)\frac{\alpha+2}{2}-(\e_{i}+\e_{j}-2)\frac{\alpha(\alpha
+3)}{12}-(\e_{j}-1)\ga(\e_{i}-\frac{\alpha}{2}-3)\frac{\alpha+2}{4}- \n
&&-\left.\left.(\e_{i}-1)(1-\bar{\ga})\frac{\alpha+2}{4}\left(\e_{j}-
\frac{\alpha}{2}-3\right)\right]g^{i}_{z}g^{j}_{\bar{z}}\right\},
\n
\beta^{k}_{z}&=&2\pi\sum_{ij}C^{k}_{ij}\left\{\left(\ga-\frac{1}{2}\right)\frac
{g^{i}g^{j}}{2}+\left[\ga+\frac{\ga(\alpha+1)}{2}-1\right]g^{i}\tilde{g}^{j}+
\right. \n
&&+\frac{\alpha+1}{2}\left[\ga\left(\frac{\alpha+1}{2}-
\frac{\e_{j}-2}{2}\right)
-\frac{\alpha-1}{4}-1\right]g^{i}g^{j}_{z\bar{z}}+ \n
&&+\left[ 1-\ga\left(\frac{\alpha+2}{2}+\frac{\e_{j}-1}{2}\right)+\frac{\alpha}
{4}\right] g^{i}g^{j}_{z}+ \n
&&+\left[\frac{\alpha+2}{4}(1-2\ga)+\frac{\e_{j}-1}{2}\ga\right] g^{i}g^{j}_
{\bar{z}}+ \n
&&+\left.\left[\frac{\alpha+1}{2}\left(\frac{\alpha-1}{4}-\ga\frac{\alpha+1}{2}
\right)+(\e_{j}-1)\ga\frac{\alpha+1}{4}+(\e_{i}-1)(\ga-1)\frac{\alpha+1}{4}
\right]g^{i}_{z}g^{j}_{\bar{z}}\right\},
\n
\beta^{k}_{\bar{z}}&=&2\pi\sum_{ij}C^{k}_{ij}\left\{\left(\bar{\ga}-\frac{1}{2}
\right)\frac
{g^{i}g^{j}}{2}+\left[\bar{\ga}+\frac{\bar{\ga}(\alpha+1)}{2}-1\right]g^{i}
\tilde{g}^{j}+\right. \n
&&+\frac{\alpha+1}{2}\left[\bar{\ga}\left(\frac{\alpha+1}{2}-\frac{\e_{j}-2}{2}
\right)-\frac{\alpha-1}{4}-1\right]g^{i}g^{j}_{z\bar{z}}+ \n
&&+\left[ 1-\bar{\ga}\left(\frac{\alpha+2}{2}+\frac{\e_{j}-1}{2}\right)+
\frac{\alpha}{4}\right] g^{i}g^{j}_{\bar{z}}+ \n
&&+\left[\frac{\alpha+2}{4}(1-2\bar{\ga})+\frac{\e_{j}-1}{2}\bar{\ga}\right]
g^{i}g^{j}_{z}+ \n
&&+\left.\left[\frac{\alpha+1}{2}\left(1+\frac{\alpha-1}{4}-\bar{\ga}
\frac{\alpha+1}{2}
\right)+(\e_{j}-1)\bar{\ga}\frac{\alpha+1}{4}+(\e_{i}-1)(\bar{\ga}-1)
\frac{\alpha+1}{4}\right]g^{i}_{z}g^{j}_{\bar{z}}\right\}, \nonumber
\eeqa
\newline
where we use the notation

\beqa
&&\e_{i}=2-h_{i}-\bar{h}_{i}, \n
&&\ga=\ga_{ijk}=\frac{h_{k}+h_{i}-h_{j}}{2h_{k}}, \n
&&\bar{\ga}=\bar{\ga}_{ijk}=\frac{\bar{h}_{k}+\bar{h}_{i}-\bar{h}_{j}}
{2\bar{h}_{k}}, \n
&&\alpha=\alpha_{ijk}=h_{k}+\bar{h}_{k}-h_{i}-\bar{h}_{i}-h_{j}-\bar{h}_{j}
 \nonumber
\eeqa

\section{Conclusions}

   We have expressed the deviations from scale invariance resulting from a
small
 perturbation of a general conformal theory by the $\beta$-functions of eq.
 (\ref{40}). These are given in terms of conformal data of an unperturbed
model, the spectrum of dimensions and the operator product coefficients. We
have
 seen that besides the primary fields the renormalization induced descendent
field and also dilatonic terms with explicit world sheet metric dependence.
These are of particular importance for non unitary theories where these terms
may still be relevant. This explicit coupling to the background metric has to
be taken into account when this metric becomes dynamical, that is when trying
to couple the perturbed theory to quantum gravity.

   The specific form of the $\beta$-functions in eq. (\ref{40}) depends of
course on the specific regularization chosen, since the precise meaning of
adding an operator to the action with a certain coupling strength depends on
the regularization scheme. For an example of such dependence, suppose that
the sharp cut off scheme of eq. (\ref{9}) is replaced by a smooth
regularization by substituting

\beq
\theta(d(z,\om)-a)~\rightarrow~\int^{\infty}_{0} d\lambda\theta(d(z,\om)-
\lambda a)f(\lambda) \label{41}
\eeq
\newline
where $f(\lambda)$ is some fixed smooth function which vanishes fast enough
when $\lambda\rightarrow 0$ and $\lambda\rightarrow\infty$. In order to
preserve
 the meaning of the coupling strengths $g^{k}$, normalize $f$ to unity

\beq
\int^{\infty}_{0} d\lambda f(\lambda)=1 \label{42}
\eeq
\newline
Repeating our analysis with this smooth regularization we see that the
contribution in (\ref{40}) to the $\beta$-function of operator of dimension
$h_{k}$ coming from the product of two perturbing operators of dimensions
$h_{i}$ and $h_{j}$ gets multiplied by the quantity

\beq
\int^{\infty}_{0} d\lambda f(\lambda)\lambda^{\e_{i}+\e_{j}-\e_{k}} \label{43}
\eeq
\newline
i.e. by the $h_{k}-h_{i}-h_{j}+2$ moment of the function $f$. In the particular
 case $\e_{i}+\e_{j}=\e_{k}$ which corresponds to a logarithmic singularity in
the integration of the location of $O_{j}$ near that of $O_{i}$, the $\beta$-
function coefficient is universal due to (\ref{42}).

\section*{Appendix}

   Here we shall calculate the geodesical distance between points $z$ and $\om$
on two dimensional surface with conformal metric $g_{\alpha\beta}=e^{\s (z)}
diag(1,1)$. The geodesical distance is

\beq
d(z,\om)=\frac{1}{\sqrt{2}}\int_{0}^{1} d\tau [g_{\alpha\beta}\dot{v}^{\alpha}
\dot{\bar{v}}^{\beta}]^{\frac{1}{2}}~=\int_{0}^{1} d\tau e^{\s (v(\tau))}
|\dot{v}(\tau)| \label{a1}
\eeq
\newline
where $\tau$ is parameter which varies from 0 to 1 along the geodesical curve
 $v(\tau)$ and dot means the derivative with respect to
 this parameter. The equation for
geodesic curve

\beq
\ddot{v}(\tau)+2\pa\s (v(\tau))\dot{v}^{2}(\tau)=0 \label{a2}
\eeq
\newline
and analogous equation for $\bar{v}(\tau)$
may be solved in perturbations over $\s$. In the zeroth order we have

\beq
v_{0}(\tau)=\tau z+(1-\tau)\om \label{a3}
\eeq
\newline
Let us call by $v_{1},v_{2}$ contributions of the first and the second
order in $\s$ to $v(\tau)$. Expansion in the Taylor series gives

\beqa
&&|\dot{v}_{0}+\dot{v}_{1}+\dot{v}_{2}|=|\dot{v}_{0}|\left( 1+
\frac{\dot{v}_{1}}{2\dot
{v}_{0}}+\frac{\dot{v}_{2}}{2\dot{v}_{0}}-\frac{\dot{v}^{2}_{1}}{8\dot{v}^{2}
_{0}}+\right.\label{a4} \\
&&+\left.\frac{|\dot{v}_{1}|^{2}}{|\dot{v}_{0}|^{2}}~+complex~conjugated\right)
\nonumber
\eeqa
\newline
Substituting this and (\ref{a3}) into expression (\ref{a1}) for $d$ we get

\beq
d(z,\om)=|z-\om|\left[ 1+\int_{0}^{1} d\tau\s (v_{0}(\tau))+\int_{0}^{1}
d\tau \left(\frac
{\dot{v}_{1}}{2(z-\om)}+c.c.\right)+\frac{1}{2}\int_{0}^{1} d\tau\s^{2}
(v_{0}(\tau))+\right. \label{a5}
\eeq
\[ +\int_{0}^{1} d\tau \left(\pa\s (v_{0}(\tau))v_{1}(\tau)+c.c.\right)
+\frac{1}{2}
\int_{0}^{1} d\tau\s (v_{0}(\tau))\left(\frac{\dot{v}_{1}}{z-\om}+c.c.\right)+
 \]
\[\left. +\frac{1}{4}\int_{0}^{1} d\tau\frac{|\dot{v}_{1}|^{2}}
{|\dot{v}_{0}|^{2}}+
\frac{1}{2}\int_{0}^{1} d\tau \left(\frac{\dot{v}_{2}}{2\dot{v}_{0}}+c.c.
\right)-
\frac{1}{8}\int_{0}^{1} d\tau \left(\frac{\dot{v}_{1}^{2}}{\dot{v}_{0}^{2}}
+c.c.\right)\right] \]
\newline
The last and before last integrals may be dropped out since last is higher
order in derivative of $\s$, and before last is total derivative over $\tau$.
 So in considered order of perturbation
we need only to know $v_{1}$, which may be found from (\ref{a2}):

\beq
v_{1}(\tau)=\pa\s (\om)(z-\om)^{2}\tau (1-\tau)+\frac{1}{3}\pa\bar{\pa}\s (\om)
|z-\om|^{2}(z-\om)\tau (1-\tau^{2}) \label{a6}
\eeq
\newline
Substituting this into (\ref{a5}) after integration over $\tau$ we get
the answer

\beq
 d(z,\om ) = |z-\om | \left( 1+\s (\om )+\frac{1}{2}(\pa\s (\om )
 (z-\om )+\bar{\pa}\s (\om )(\bar{z}-\bar{\om })) + \frac{1}{3}
 \pa\bar{\pa}\s (\om )|z-\om |^{2}~+\right. \label{dap}
\eeq
\[+\frac{1}{2}\s ^{2}(
 \om )+\frac{1}{2}\s (\om )\pa\s (\om )(z-\om ) +
 \frac{1}{2}\s (\om )\bar{\pa}\s (\om )(\bar{z}-\bar{\om }) + \]
\[\left.+\frac{1}{3}\s (\om )\pa\bar{\pa}\s (\om )|z-\om |^{2}
 +\frac{1}{4}\pa\s (\om )\bar{\pa}\s (\om )|z-\om |^{2}\right)= \]
\[ = e^{\s (\om)}|z-\om | \left( 1+\frac{1}{2}(\pa\s (\om )
 (z-\om )+\bar{\pa}\s (\om )(\bar{z}-\bar{\om })) + \frac{1}{3}
 \pa\bar{\pa}\s (\om )|z-\om |^{2}~+\right. \]
\[\left. +\frac{1}{4}\pa\s (\om )\bar{\pa}\s (\om )|z-\om |^{2})\right) \]
\newline
where $y=|z-\om|,\theta=arg(z-\om)$ and
we collected all powers of $\s$ into
exponential in the considered second order of perturbation in $\s$.

\section*{Acknowledgements}

                ~We thank M.Dine, A.Forge, D.Kutasov, E.Rabinovici and
A.Shwimmer and A.Tseytlin for useful discussions.
 S.E. would like to thank the International
School for Advanced Studies, SISSA, Trieste, for hospitality while part of this
 work was done.
{}~

\end{document}